\documentclass[final,3p,times,twocolumn]{elsarticle}

\usepackage{graphicx}

\usepackage{amsmath}
\usepackage{amssymb}

\usepackage{lineno}

\usepackage{color}
\usepackage[usenames,dvipsnames]{xcolor}

\usepackage{ulem}

\newcommand{\res}[1]{{\mathcal{R}_{#1}}}

\journal{Nuclear Instruments and Methods in Physics Research Section A}

\begin{document}

\begin{frontmatter}




\title{Event plane resolution correction for azimuthal anisotropy in wide centrality bins}


\author{Hiroshi Masui}
 \ead{hirmasui@gmail.com}
\author{Alexander Schmah}
 \ead{ASchmah@lbl.gov}
\author{A.M. Poskanzer}
 \ead{AMPoskanzer@lbl.gov}
 \address{
  Lawrence Berkeley National Laboratory,
  Berkeley, CA 94720, USA
}

\begin{abstract}
We provide a method to correct the observed azimuthal anisotropy
in heavy-ion collisions for the event plane resolution in a wide centrality bin.
This new procedure is especially useful for rare particles,
such as $\Omega$ baryons and $J/\psi$ mesons, which are difficult to measure in
small intervals of centrality.
Based on a Monte Carlo calculation with simulated $v_2$ and multiplicity,
we show that some of the commonly used methods have a bias 
of up to 15\%.
\end{abstract}

\end{frontmatter}

\section{Introduction}
\label{sec:introduction}

  Azimuthal anisotropy is one of the key observables to study the properties of matter created
  in high energy heavy-ion collisions~(see e.g.~\cite{Voloshin:2008dg}).
  It is characterized by the Fourier decomposition of the azimuthal particle distribution with respect to the participant
  plane~\cite{Voloshin:1994mz}

  \begin{linenomath}
  \begin{equation}
    \frac{dN}{d\phi} = \frac{N_0}{2\pi}
    \left(1 + 2\sum_{n=1}^{\infty}v_n\cos{[n(\phi-\Psi_{\rm PP_n})]}
	\right),
    \label{eq:PP}
  \end{equation}
  \end{linenomath}
  where $N_0$ is the number of particles in the event, $v_n$ is the $n$-th harmonic coefficient,
  $\phi$ is the azimuthal angle of particles and $\Psi_{{\rm PP}_n}$
  is the azimuthal angle of the $n^{th}$ harmonic participant plane. The participant plane is the
  symmary plane of the participants. 

  One of the standard methods is to extract $v_n$ by using the reconstructed
  event plane from the detected participant particles~\cite{Poskanzer:1998yz}.
  The most important task in this method is to estimate the participant plane from the measured particles
  for each harmonic $n$.
  The estimated participant plane is defined as the {\it event plane} $\Psi_n$~($-\pi/n \le \Psi_n < \pi/n$),
  but due to the finite multiplicity in nuclear collisions and fluctuations, the event plane can be different from the
  participant plane.
  The observed $v_{n}^{\rm obs}(M)$ for a given, small, centrality range $M$ 
  must be corrected by the {\it event plane resolution} $\res{n}(M)$
  in order to take into account the difference between true participant plane and event plane
 
   \begin{linenomath}
  \begin{equation}
    v_n(M) = \frac{\left<\left<\cos{[n(\phi-\Psi_m)]}\right>\right>_M}
    {\left<\cos{[n(\Psi_m-\Psi_{{\rm PP}_{n}})]}\right>_M}
    \equiv \frac{v_n^{\rm obs}(M)}{\res{n}(M)},
    \label{eq:vn}
  \end{equation}
  \end{linenomath}
  where $m$ is the harmonic of the event plane and $n = km$ is the harmonic of interest.
  The integer $k$ is taken as unity in this paper.
  Brackets denote the average over events, while double brackets denote the average over particles in all events.
  Subscript $M$ on the bracket emphasizes that the average is taken for a
  given centrality $M$.
  For simplicity, we will omit $M$ from the observables, for example, we will
  write $N_0$ instead of $N_0(M)$.
  In experiments the participant plane angle $\Psi_{{\rm PP}_n}$ in Eq.~(\ref{eq:PP}) is not known.
  Therefore at least two subevent planes are necessary in order to calculate the
  event plane resolution~\cite{Poskanzer:1998yz}.
  An average $v_{n}$ over a wider centrality range $R$ can be calculated once
  $v_{n}^{\rm obs}$ and $\res{n}$ are determined within the range $R$

  \begin{linenomath}
  \begin{equation}
  \frac{
    \displaystyle{\int_R dM ~ N_0 \frac{v_n^{\rm obs}}{\res{n}}}
  }{
    \displaystyle{\int_R dM ~ N_0}
  }
   \equiv \left<\frac{v_n^{\rm obs}}{\res{n}}\right> = \left<v_n\right>.
    \label{eq:vnR}
  \end{equation}
  \end{linenomath}
  We introduced brackets $\left<...\right>$ for simplicity, which
  denote the average over a wide centrality range weighted 
  by particle multiplicity $N_0$ for a given phase space
  (e.g. for a given transverse momentum range).

  $\res{n}$ depends on multiplicity and $v_n$ itself,
  therefore Eq.~(\ref{eq:vnR}) requires that $v_n^{\rm obs}$ and $\res{n}$
  are measured in sufficiently small centrality intervals.
  However, for rare particles (e.g. $\Omega$, $J/\psi$) it is not always possible to do so.
  One of the conventional approaches (e.g. see Ref.~\cite{Aggarwal:2010mt})
  is to average $v_n^{\rm obs}$ as well as $\res{n}$ separately in a wide
  centrality range, weighted by the corresponding particle yields,
  and then divide $\left<v_n^{\rm obs}\right>$ by $\left<\res{n}\right>$.
  However, this approach systematically overestimates $v_n$ as we will
  discuss in Section~\ref{sec:simulation}. The main point of this paper
  is the following inequality

  \begin{linenomath}
  \begin{equation}
    \left<\frac{v_n^{\rm obs}}{\res{n}}\right>
      \neq \frac{\left<v_n^{\rm obs}\right>}{\left<\res{n}\right>}
      \neq \left<v_n^{\rm obs}\right> \left<\frac{1}{\res{n}}\right>
  \end{equation}
  \end{linenomath}
  where the left hand side is the correct average over a wide centrality bin,
  while the right hand side shows commonly used approximations. However, it is easy
  to avoid these approximations.

  We will show the proper way to correct for the finite event plane resolution in wide centrality bins. Eventhough
  the resolution is an event-by-event quantity, the calculation of the sum over individual particles for the
  flow coefficients must be done with the inverse of the reolution for the proper centrality.
  In Section~\ref{sec:method}, we derive the equations used to calculate $\left<v_n\right>$ in our approach.
  We also show that the derived equations are equivalent to the average calculated from narrow centrality bins
  (see Eq.~(\ref{eq:vnR})).
  In Section~\ref{sec:simulation}, we show a simple Monte Carlo simulation to demonstrate the validity
  of the method,
  for the case of $v_{2}$. Based on a preliminary version of this paper~\cite{Masui:2012zh}, the method has
  been appled allready in several publications~\cite{Bairathi:2015uba,Lomnitz:2016rpz,Nasim:2014iea}.

\section{Implementation}
\label{sec:method}

  We now show two practical implementations to correct $v_{n}$ for the event plane resolution in wide centrality bins.
  There are two or three steps to calculate $v_n$ in wide centrality bins
  \begin{enumerate}
    \item Determine the event plane resolution $\res{n}$ as a function of $M$ in narrow centrality ranges.
    \item Analyse $v_n$ with the weights $1/\res{n}$ for any, wide, centrality range
    of interest.
    \item In addition for the event-plane method, one must multiply the result by the average weight.
  \end{enumerate}
  In the following subsections, we discuss detailed implementations of how to apply corrections
  for different types of particle identification.
  For the sake of simplicity, we assume that non-flow effects are negligible,
  and all correlations between particles are induced by flow. A systematic study of other effects can be found
  in Ref.~\cite{Nasim:2014iea}.
  
  The azimuthal particle distribution with respect to the event plane can be written as
  \begin{linenomath}
  \begin{equation}
    \frac{dN}{d(\phi-\Psi_m)} = \frac{N_0}{2\pi}
    \left(1 + 2\sum_n^{\infty}v_n^{\rm obs}\cos{[n(\phi-\Psi_m)]}\right).
  \end{equation}
  \end{linenomath}

  \subsection{Event-by-event particle identification}
  \label{subset:ebye_pid}

  If the particle of interest can be identified on an event-by-event basis,
  one can directly calculate $\cos{[n(\phi-\Psi_m)]}$ for every particle,
  corrected with the event plane resolution for the corresponding centrality $M$
  
  \begin{linenomath}
  \begin{equation}
   \frac{ \cos{[n(\phi-\Psi_m)]}}{\res{n}},
    \label{eq:non-event}
  \end{equation}
  \end{linenomath}
  where $\res{n}(M)$~is supposed to be averaged over many events in advance.
  The event and centrality average in the range of multiplicities $R$ of term~(\ref{eq:non-event}) over $\phi-\Psi_m$
  reduces in this case to Eq.~(\ref{eq:vnR}):
  \begin{linenomath}
  \begin{align}
  & \frac{
     \displaystyle{
       \int_R dM \int_0^{2\pi} d(\phi-\Psi_m) \frac{dN}{d(\phi-\Psi_m)}
       \frac{\cos{[n(\phi-\Psi_m)]}}{\res{n}}
     }
   }{
     \displaystyle{
       \int_R dM \int_0^{2\pi} d(\phi-\Psi_m) \frac{dN}{d(\phi-\Psi_m)}
     }
   }  \nonumber \\
   & = \frac{
    \displaystyle{\int_R dM ~ N_0 \frac{v_n^{\rm obs}}{\res{n}}}
  }{
    \displaystyle{\int_R dM ~ N_0}
  }
  = \left<v_n\right>
  \label{eq:vn_ebye}
  \end{align}
  \end{linenomath}
  The main difference of our implementation to the
  conventional approach is the event-by-event resolution correction in
  Eq.~(\ref{eq:vn_ebye}). As we have mentioned earlier, event-by-event $\res{n}$
  correction does not mean that the $\res{n}$ should be calculated
  event-by-event, but rather that $\res{n}(M)$ is calculated for many events and then the resolution
  correction is made event-by-event as a function of centrality, $M$.

  \subsection{Statistical particle identification}

  There are two approaches in case the particle yield of interest can only be extracted statistically:
  the invariant mass fit method and the event plane method. The invariant mass
  fit method is almost equivalent to the one introduced in Section~\ref{subset:ebye_pid},
  while the event plane method needs one additional step to obtain the final
  $v_{n}$.

  \subsubsection{Invariant mass fit method}

  The invariant mass method~\cite{Borghini:2004ra} is quite useful to analyse particles that are
  detected through their decay products,
  such as $K^0_s \rightarrow \pi^+\pi^-$, $\Lambda \rightarrow p\pi^-$ and so on.
  The point of this method is to calculate the mean azimuthal angle relative to the event plane,
  $\cos{[n(\phi-\Psi_m)]}$, as a function of invariant mass $M_{\rm inv}$

  \begin{linenomath}
  \begin{align}
    v_{n}^{S+B}(M_{\rm inv}) & = \left\langle \cos\left[n(\phi - \Psi_{m})
      \right]_{{\rm inv}} \right\rangle, \nonumber \\
    & = v_{n}^{S} \frac{S}{S+B}(M_{\rm inv})  
     + v_{n}^{B}(M_{\rm inv})\frac{B}{S+B}(M_{\rm inv}). \nonumber \\
    \label{eq:finvmass_v2_fit}
  \end{align}
  \end{linenomath}  
  where $S$ is the signal yield peaked at the mass of the partcle, $B$ is the smooth background yield,
  $v_n^S$, $v_n^B$ and $v_n^{S+B}$ are the $v_n$ for signal, background and
  total particles, respectively.
  Signal and background contributions are decomposed by taking into account the measured signal-to-background
  ratio and using a parametrization of the background $v_{n}^{B}$ shape
  by assuming that $B(M_{\rm inv})$, and $v_n^B(M_{\rm inv})B(M_{\rm inv})$
  are smooth functions of $M_{\rm inv}$~\cite{Borghini:2004ra}, while $v_n^SS(M_{\rm inv})$ is peaked at
  $M_{\rm inv}$.
  Since the average cosine is calculated in this approach, one can directly
  extract the $v_{n}^{S}$ by subtracting the background contribution.
  The only modification is to add a weight $1/\res{n}$~on an event-by-event basis
  when one fills the histograms for $\cos{[n(\phi-\Psi_m)]}$ versus invariant mass
  similar to Eq.~(\ref{eq:vn_ebye}).

  \subsubsection{Event plane method}
  \label{subsec:eventplanemethod}

  The event plane method~\cite{Poskanzer:1998yz} with $\phi-\Psi_m$ binning is an alternative approach
  to a $v_n$ analysis when
  particle yields can be determined only statistically after a background is subtracted.
  This method requires the decomposition of signal and background in several
  $\phi-\Psi_m$ bins (typically 10-20), while the invariant mass fit method only
  requires the decomposition once. The event plane method does not need the
  assumption of a background $v_n$ shape. Therefore this method 
  has some advantage, especially when the signal to background ratio is poor.
  
  The first step in the event plane method is the signal extraction for a given
  $\phi-\Psi_m$ bin
  \begin{linenomath}
  \begin{align}
    N^{\res{}}(\phi-\Psi_m) & =
    \int_R dM ~ \frac{1}{\res{n}} \frac{dN}{d(\phi-\Psi_m)},
    \label{eq:n_phipsi}
  \end{align}
  \end{linenomath}
  where $N^{\res{}}(\phi-\Psi_m)$ is the number of particles for a given $\phi-\Psi_m$
  bin, weighted for each centrality bin with the inverse of the event plane resolution.
  The difference to the conventional method is the weight $1/\res{n}$ on
  the particle yields which will properly take into account the centrality
  dependence of the event plane resolution.
  Second, one integrates over $\phi-\Psi_m$ to
  calculate $v_n$
  \begin{linenomath}
  \begin{align}
 & \frac{
      \displaystyle{
	\int_0^{2\pi} d(\phi-\Psi_m) N^{\res{}}(\phi-\Psi_m) \cos{[n(\phi-\Psi_m)]}
      }
    }{
      \displaystyle{
	\int_0^{2\pi} d(\phi-\Psi_m) N^{\res{}}(\phi-\Psi_m)
      }
    } \nonumber
   \end{align}
   \end{linenomath}
     and then using Eq.~(\ref{eq:n_phipsi}) gets
  \begin{linenomath}
  \begin{align}
    \label{eq:stat_v2_phipsi_integration}
 	\left<v_{n}^{\res{}}\right> =  \\
  & \frac{
     \displaystyle{
       \int_R dM \int_0^{2\pi} d(\phi-\Psi_m) \frac{dN}{d(\phi-\Psi_m)}
       \frac{\cos{[n(\phi-\Psi_m)]}}{\res{n}}
     }
   }{
     \displaystyle{
     \int_R dM \int_0^{2\pi} d(\phi-\Psi_m) \frac{dN}{d(\phi-\Psi_m)} \frac{1}{\res{n}}
     }}. \nonumber
   \end{align}
   \end{linenomath}
  Note that this equation includes the $1/\res{n}$ weights
  on the particle yields in the denominator.
  One could immediately notice that the numerator is identical to
  Eq.~(\ref{eq:vn_ebye}) with the additional normalization
  $1/\left(\int_R dM N_0\right)$. 
  The denominator becomes
  \begin{linenomath}
  \begin{align}
    &  \int_R dM \frac{N_0}{\res{n}}
       =  \left(\int_R dM N_0\right) \left<\frac{1}{\res{n}}\right>. \nonumber
  \end{align}
  \end{linenomath}
  The normalization factor $\int_R dM N_0$ is
  cancelled out between numerator and denominator.
  Thus $\left<v_n\right>$ is obtained from Eq.~(\ref{eq:stat_v2_phipsi_integration}) by
  multiplying by the average weight, $\left<1/\res{n}\right>$:
   \begin{linenomath}
   \begin{align}
    \left<v_n\right>  =  \left<v_n^{\res{}}\right> \left<\frac{1}{\res{n}}\right>.
    \label{eq:vn}
   \end{align}
  \end{linenomath} 
  
  We would like to emphasize two important points for the event plane method.
  First, one must take an average of the inverse event
  plane resolution $1/\res{n}$ (not an average of $\res{n}$).
  Second, the integral $\int_R dM$ in Eq.~(\ref{eq:stat_v2_phipsi_integration}) for both numerator
  and denominator should be calculated in the same phase space.
  For instance, if one measures the $v_n$ as a
  function of transverse momentum $p_T$, one should calculate $\left<1/\res{n}\right>$
  for each $p_T$ bin.

  This new implementation can also be applied for the scalar product
  method~\cite{Adler:2002pu, Luzum:2012da}. The scalar product method is
  an extension of the methods described in this paper where in addition to the event plane angle the magnitude of the
  Q-vector~\cite{Poskanzer:1998yz} is considered for both $v_n^{\rm obs}$ and $\res{n}$.

\section{Simulation results}
\label{sec:simulation}

  In this section we validate our implementation by a simple Monte Carlo
  simulation for $v_2$($p_T$). As we have mentioned earlier, this approach can
  be applied for any harmonic of interest.

  \begin{figure}[htbp]
    \includegraphics[width=0.45\textwidth]{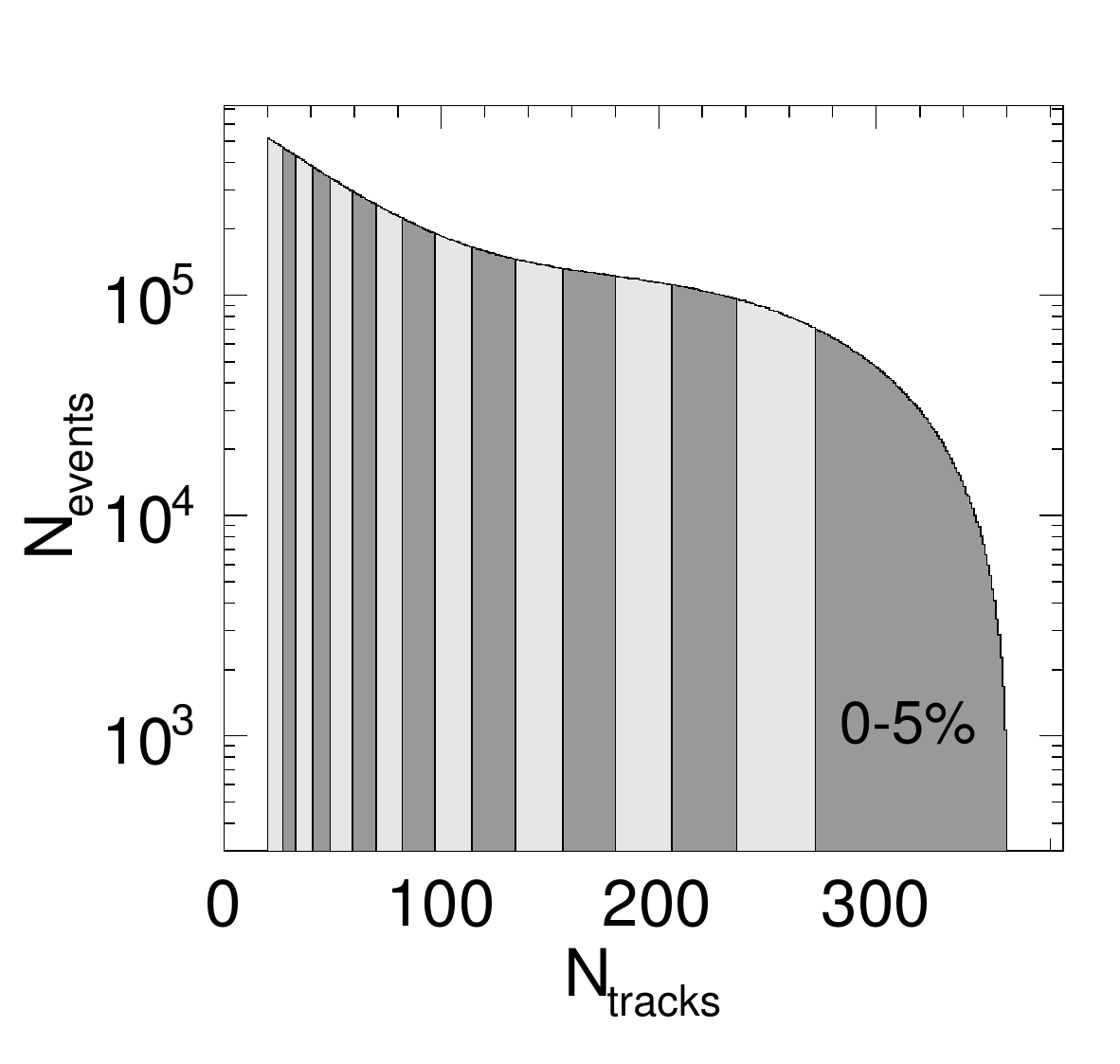}
    \caption{
      \label{fig:multiplicity}
      Input multiplicity distribution in the simple Monte Carlo simulation. The
      shaded area show 0-5\%, 5-10\%, ..., 75-80\% centrality classes.
    }
  \end{figure}

  Figure~\ref{fig:multiplicity} shows the input multiplicity distribution in
  0-80\% centrality for the 50 million generated events.  The event centrality classes are determined by using the
  multiplicity distribution as it is typically done in the experimental data.
  We divide the multiplicity distribution into 16 bins with 5\% increments within 0-80\%.

\begin{figure}[htbp]
    \includegraphics[width=0.45\textwidth]{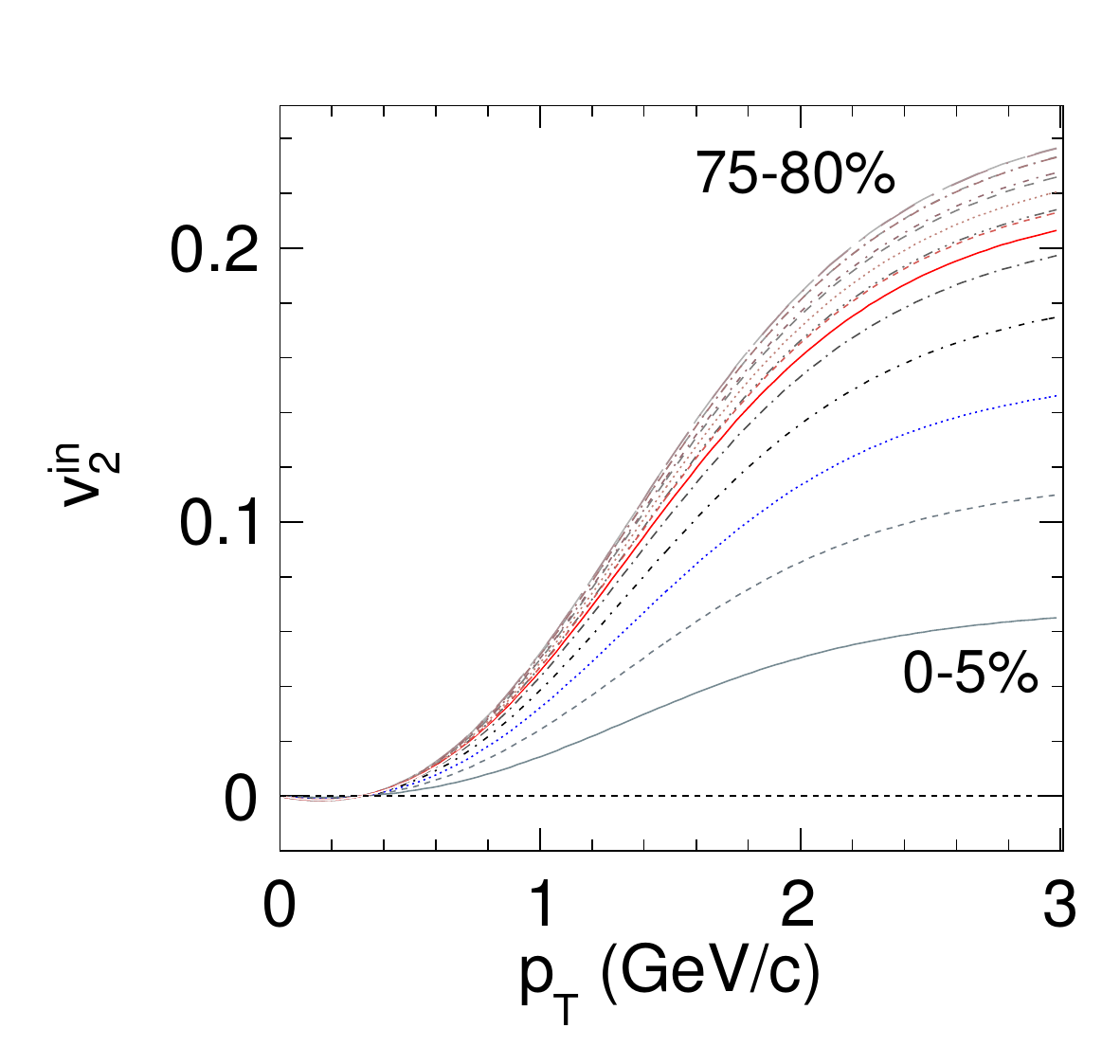}
    \caption{
      \label{fig:v2cent}
      (Color online)
      Input $v_{2}(p_{T})$ distributions for different centralities.
      Each line corresponds to a 5\% centrality range.
      $v_{2}^{\rm in}(p_{T})$ increases towards more peripheral events. The applied Gaussian smearing is not shown.
    }
  \end{figure}

  For every event a number of tracks according to the input multiplicity distribution was sampled.
  A Boltzmann like $p_{T}$ distribution from 0.25--3 GeV/$c$ was used.
  The input $v_{2}^{\rm in}(p_{T})$ for every centrality bin
  was fixed to an empirical parametrization of the following form 
  \begin{linenomath}
  \begin{align}
     v_{2}^{\rm in}(p_{T}) & = A(1-e^{-p_{T}}) \left( \frac{a}{1+e^{-(p_{T}-b)/c}} -d \right)
    \label{eq:v2in}
  \end{align}
  \end{linenomath}
  which describes the shape of observed $v_2(p_T)$ in heavy ion collisions.
  The parameter $A$ was increased towards more peripheral centralities. Furthermore we have added for every event a
  Gaussian smearing on $v_{2}^{\rm in}$ with a width of 0.05.
  Figure~\ref{fig:v2cent} depicts the used $v_{2}^{\rm in}(p_{T})$ for the 16 
  centrality bins. The event Q-vector was calculated according to Ref.~\cite{Poskanzer:1998yz}. 

  \begin{figure}[htbp]
    \includegraphics[width=0.45\textwidth]{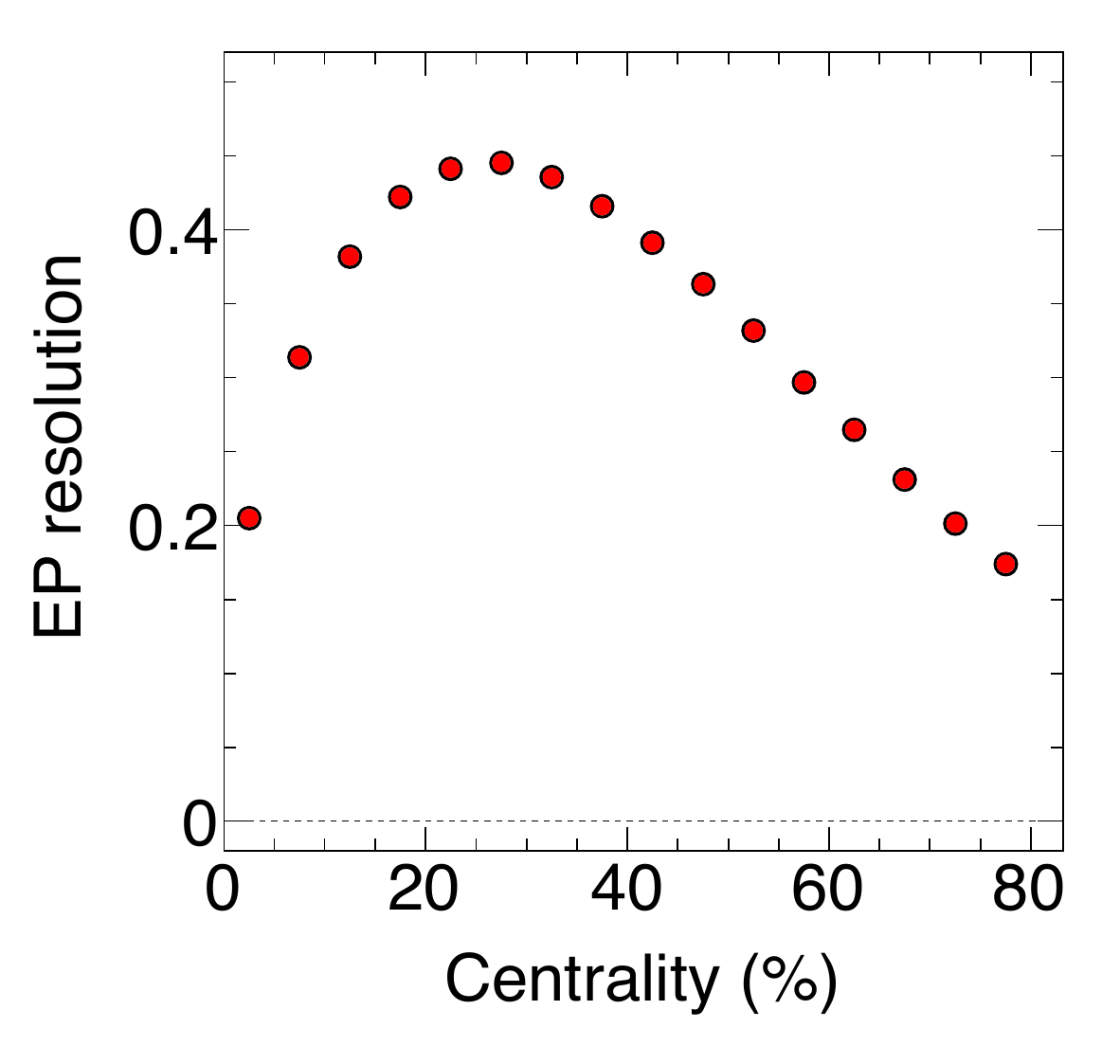}
    \caption{
      \label{fig:resolution}
      (Color online)
      Subevent-plane resolution for the second harmonic as a function of
      centrality. The resolution is calculated from random subevents.
    }
  \end{figure}
  
  Figure~\ref{fig:resolution} shows the random subevent plane resolution as a function of
  centrality. For each event, particles are randomly divided into two different
  groups in order to evaluate the resolution $\res{2}$. The result shown
  here is the subevent plane resolution by using Eq.~(14)
  in Ref.~\cite{Poskanzer:1998yz}.
  The resolution reaches a maximum of 0.45 around 30\% centrality, it decreases towards
  more central and peripheral events because of a lower $v_2$ in central and
  less multiplicity in peripheral events. Then the full event plane resolution, which is used in this paper, 
  is calculated from Eq.~(11) in Ref.~\cite{Poskanzer:1998yz}.

  \begin{figure}[htbp]
    \includegraphics[width=0.45\textwidth]{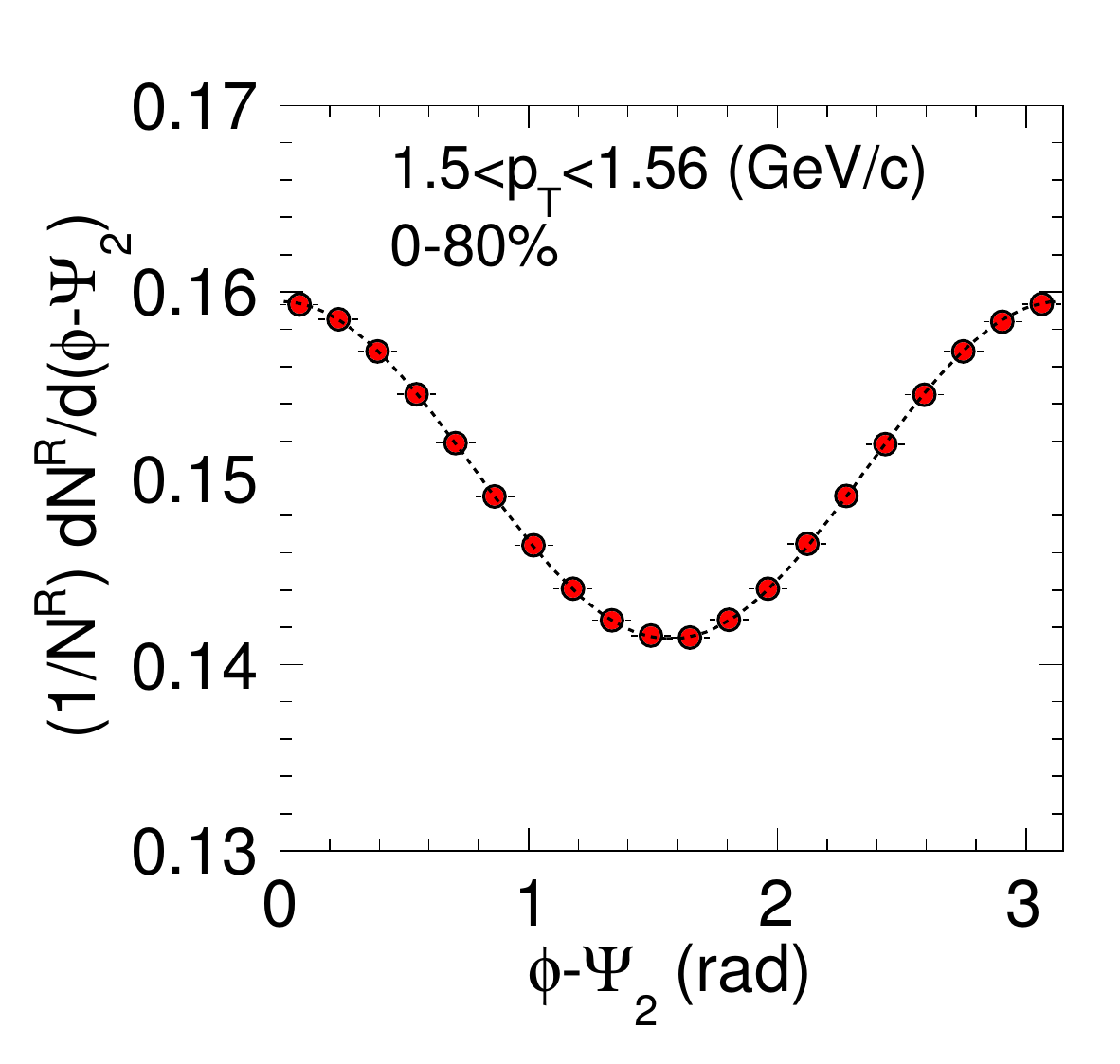}
    \caption{
      \label{fig:psi_phi}
      (Color online)
      Azimuthal distribution of particles with respect to the second harmonic
      event plane at 1.5~$< p_T <$~1.56~GeV/$c$ in 0--80\% centrality bin. The
      dashed line is the fit result for
      1+2$v_2^{\rm obs}\cos{[2\phi-2\Psi_2]}$.  }
  \end{figure}

  Figure~\ref{fig:psi_phi} shows an example of a particle azimuthal distribution with
  respect to the second harmonic event plane in a narrow $p_T$ bin. The particle
  yields are weighted by the inverse event plane resolution as indicated by $N^{\res{}}$
  in the y-axis title.
  The dashed line represents a fit with $1+2v_2^{\rm obs, \res{}}\cos{[2\phi-2\Psi_2]}$.
  The yield extraction and fit are repeated for all other $p_T$ bins.

  \begin{figure}[htbp]
  \includegraphics[width=0.45\textwidth]{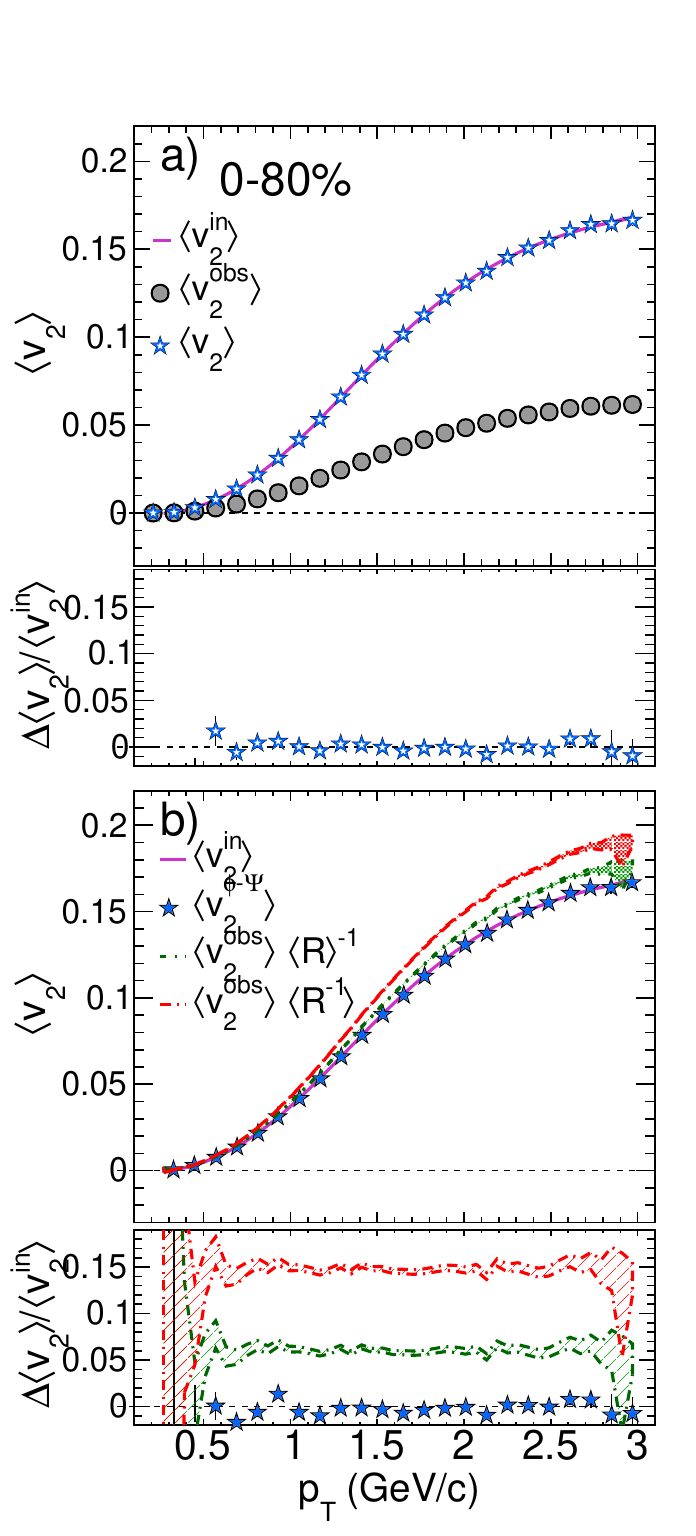}    
    \caption{
      \label{fig:v2}
      (Color online)
      Results of $v_2$ as a function of $p_T$ with different approaches of
      resolution corrections. The magenta solid line corresponds to the input $\left<v_2\right>$, gray filled circles
      are the observed $\left<v_{2}^{\rm obs}\right>$ without resolution correction, blue open stars in panel (a) are
      from the direct cosine calculation from Eq.~(\ref{eq:vn_ebye}), blue filled stars in panel (b)       
      are from the event plane method from  Eq.~(\ref{eq:vn}). Other
      lines show the observed $\left<v_{2}^{\rm obs}\right>$ with different approximate
      mean event plane resolution corrections.
     }
  \end{figure}

  Figure~\ref{fig:v2} shows $\left<v_2\right>$ as a function of $p_T$ in the 0-80\%
  centrality bin. For comparison, we also plot the observed $\left<v_{2}^{\rm obs}\right>$ without resolution
  correction as shown by solid grey circles. The difference between observed and
  corrected $\left<v_2\right>$ gives the size of the resolution correction. The input $\left<v_{2}^{\rm in}\right>$ of the
  simulation is shown as a magenta solid line.
  We tested both, the direct cosine calculation from Eq.~(\ref{eq:vn_ebye}) and the event
  plane method from Eq.~(\ref{eq:vn}),
  as shown by open blue stars and solid blue stars, respectively in panel (a) and (b).
  One can see that both direct cosine and event plane methods are consistent with the
  input $\left<v_{2}^{\rm in}\right>$ within 0.3\%.
  Panel (b) shows the observed $\left<v_{2}^{\rm obs}\right>$, corrected by the inverse mean event plane
  resolution $\left< \res{} \right>^{-1}$ (green line)
  and the mean inverse event plane resolution $\left< \res{}^{-1} \right>$ (red line).
  The relative difference $\Delta \left< v_{2} \right>$ of the corrected $\left<v_{2}\right>$
  values to the input $\left<v_{2}^{\rm in}\right>$ is shown in the lower panels.
  For the correction with $\left< \res{} \right>^{-1}$ and $\left< \res{}^{-1} \right>$ 
  we get a $\Delta \left< v_{2} \right>/\left< v_{2}^{\rm in} \right>$ of 7\% and 15\%, respectively.
  If there are enough data the resolution correction should be done
  separately for each $p_T$ bin.
\section{Conclusions}

  We have introduced an implementation to avoid event plane resolution effects of wide
  centrality bins on $v_n$ measurements. 
  In this new approach an inverse event plane resolution weight as a function of multiplicity is added 
  on an event-by-event basis to the cosine term or to the particle yields.
  We confirmed that our approach reproduces the input $\left< v_{2}^{\rm in}\right>$($p_T$) in a
  wide centrality bin, whereas the conventional correction using the mean resolution overestimates the 
  elliptic flow coefficient by 7\% for minimum bias collisions. Surprisingly, usng the mean inverse resolution,
  which was thought to be a better approximatiion, overestimates the result by twice as much.

\section*{Acknowledgements}

We thank Jean-Yves Ollitrault, Shusu Shi, and Xu Sun for discussions.
This paper was funded in part by the US Department of Energy under Contract No. DE-AC03-76SF00098.








\end{document}